\documentclass[letterpaper, 10 pt, conference]{ieeeconf}  

\IEEEoverridecommandlockouts                              

\usepackage{caption}
\usepackage{graphicx}
\usepackage{float} 
\usepackage{subcaption}
\usepackage{amsmath,amssymb,amsfonts}
\usepackage{xcolor}
\usepackage{color}
\usepackage{booktabs}
\usepackage[switch]{lineno}
\usepackage{marvosym}
\usepackage{textcomp}
\usepackage{verbatim}
\usepackage{hyperref}
\usepackage{array}
\newcounter{parnumcounter} 

\title{\LARGE \bf
Make Your AUV Adaptive: An Environment-Aware Reinforcement Learning Framework For Underwater Tasks
}

\author{Yimian Ding$^{1,\dag}$, Jingzehua Xu$^{1,\dag,}\textsuperscript{\Letter}$, Guanwen Xie$^{1}$, Shuai Zhang$^{2}$ and Yi Li$^{1}
\thanks{\dag These authors contributed equally to this work.}
\thanks{\textsuperscript{\Letter} Corresponding author.}
\thanks{$^{1}$$\,$Y. Ding, J. Xu, G. Xie and Y. Li are with the Tsinghua Shenzhen International Graduate School, Tsinghua University, Shenzhen, 518055, China. E-mail: $\{$dingym24, xjzh23, xgw24$ \}$@mails.tsinghua.edu.cn, liyi@sz.tsinghua.edu.cn.}
\thanks{$^{2}$$\,$S. Zhang is with Department of Data Science, New Jersey Institute of Technology, NJ 07102, USA. E-mail: sz457@njit.edu.}$
}

\begin{document}

\maketitle
\thispagestyle{empty}
\pagestyle{empty}

\begin{abstract}

This study presents a novel environment-aware reinforcement learning (RL) framework designed to augment the operational capabilities of autonomous underwater vehicles (AUVs) in underwater environments. Departing from traditional RL architectures, the proposed framework integrates an environment-aware network module that dynamically captures flow field data, effectively embedding this critical environmental information into the state space. This integration facilitates real-time environmental adaptation, significantly enhancing the AUV's situational awareness and decision-making capabilities. Furthermore, the framework incorporates AUV structure characteristics into the optimization process, employing a large language model (LLM)-based iterative refinement mechanism that leverages both environmental conditions and training outcomes to optimize task performance. Comprehensive experimental evaluations demonstrate the framework's superior performance, robustness and adaptability.

\end{abstract}

\section{INTRODUCTION}


AUVs are widely utilized in the marine domain due to their flexibility and practicality in areas such as resource exploration \cite{001}, ecological monitoring \cite{002}, and underwater rescue \cite{003}. However, performing more precise tasks in complex ocean environments, such as trajectory tracking and data collection \cite{004}, demands a more robust and adaptive AUV control method and poses higher requirements on AUV’s maneuverability \cite{005}. In recent years, the emergence of Reinforcement Learning (RL) has somewhat addressed these issues. RL trains robots to interact extensively with environment by setting appropriate reward functions, enabling policy improvement to achieve expert-level autonomous decision-making capabilities \cite{006}. Using RL for the control of complex robotic systems has shown promising results in various applications, including drones \cite{007, 008, 009}, legged robots \cite{010, 011, 012}, and also underwater vehicles \cite{013}. 

However, applying RL to marine robots, such as AUVs, often results in suboptimal performance \cite{014}. This is primarily because AUVs operate in ocean environments that vary in space and time, making it difficult for control methods and structural designs to adapt to unpredictable changes in fluid dynamics, such as randomly occurring turbulence fields and ocean currents \cite{015}. The hydrodynamic forces experienced by AUVs differ when they operate in deep water, near the seabed, or close to dynamic objects \cite{005}. Small changes in payload or environment during different tasks can significantly impact hydrodynamics, such as maneuverability in rotation and translation, which are difficult to model and measure \cite{016}. When using traditional RL or classical control methods like PID controllers, this can lead to a time-consuming tuning process, where adjustments must be made based on the environment or task to support new AUV configurations \cite{017}. Therefore, there is a need for a more general and adaptive RL training method for AUVs to support more complex field deployments.

\begin{figure}
\centering
\includegraphics[width=0.828\linewidth]{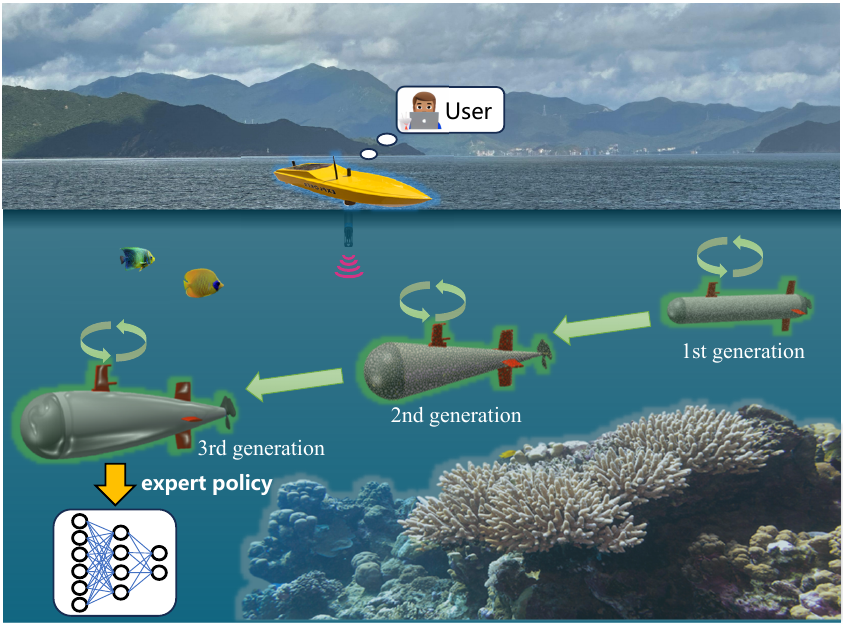}
 \caption{\small \textbf{Illustration of AUV conducting underwater tasks using our proposed environment-aware RL framework}. We jointly optimize the structure and control policy of the AUV through RL training integrated with its onboard environment-aware module, thereby enhancing its adaptivity in the complex ocean environment.\vspace{-2.0em}}
\label{fig:1}
\end{figure}

Based on the above analysis, in this study we propose a new RL framework named environment-aware RL for underwater tasks. Our main contributions can be summarized as follows:

\begin{itemize}
\item
In addition to the original value network and policy network in RL, we add a Physics-Informed Neural Networks (PINN)-enabled environment-aware network module to capture the flow field information which is incorporated into the state space. Through iterations of RL training, enabling AUV the perceptive ability and adaptivity to the complex environment.

\item
Subsequently, we design a large language model (LLM)-based optimization module in our framework. By providing the domain expertise and obtained RL training process with human feedback to the LLM for in-depth analysis, both the control strategy and the AUV’s structure can be adaptively optimized according to different environments and task requirements. Therefore, our proposed framework extends traditional RL approaches by introducing a joint optimization paradigm that co-advances strategy learning and structural design through environmental awareness.

\item

Comprehensive simulations demonstrate that the proposed framework enhances expert-level decision-making capabilities for AUVs while effectively optimizing structural design, thereby significantly improving environmental adaptability. This ultimately achieves superior performance compared to traditional RL frameworks.

\end{itemize}

\section{RELATED WORK}

Numerous previous studies have applied RL algorithms to train AUVs to complete underwater tasks. Tang \textit{et al.} \cite{4} implemented a Deep Q-Network (DQN) approach for dynamic AUV navigation to optimize data collection efficiency \cite{9069205}. Ding \textit{et al.} \cite{6} developed a parameter prediction-based RL algorithm to address suboptimal policies convergence in multi-AUV task allocation under ocean currents. While these works demonstrate RL's foundational potential in underwater operations, practical implementation challenges persist \cite{7}. Under extreme disturbances, conventional RL approaches may fail to achieve theoretical optimality due to inherent limitations in AUV hydrodynamics and maneuverability \cite{8}.


Building on this foundation, several studies have enhanced RL algorithms for complex underwater environment. Sun \textit{et al}. \cite{10} improved the deep deterministic policy gradient (DDPG) algorithm by incorporating optimized sampling and motion evaluation for AUV path-following in interference-prone environments. Xu \textit{et al}. \cite{11} reformulated the underwater pursuit-evasion (UPE) game \cite{12} as a finite-horizon Markov game and introduced a decentralized offline RL framework, achieving high performance in varying UPE scenarios. While these studies somehow improve AUV's adaptivity in complex underwater environment, none address the AUV’s environmental adaptability or the impact of its boundary structure on training. The environment-aware RL framework proposed here effectively resolves these issues.

\section{METHODS}


This section details the implementation of the proposed environment-aware RL framework. A schematic diagram of the framework's overall structure is provided in Fig. 2.

\subsection{The Environment-Aware Module}

In the proposed RL framework, the environment-aware module is responsible for introducing underwater flow field information into the RL training process, thereby enabling the AUV to perceive its surrounding environment during task execution. To better integrate this module into the RL algorithm, similar to the policy network and value network, we construct this module using a PINN \cite{RAISSI2019686} and name it the environment-aware network $\mathcal{U}_E(\mathbf{\vec{x}},t;\theta)$, where $\theta$ is the network's parameter. Specifically, the network can output the corresponding flow field velocity $\mathbf{\vec{v}}$ and pressure $p$ based on the input position $\mathbf{\vec{x}}$ and time $t$. 

\begin{figure*}
\centering
\includegraphics[width=0.928\linewidth]{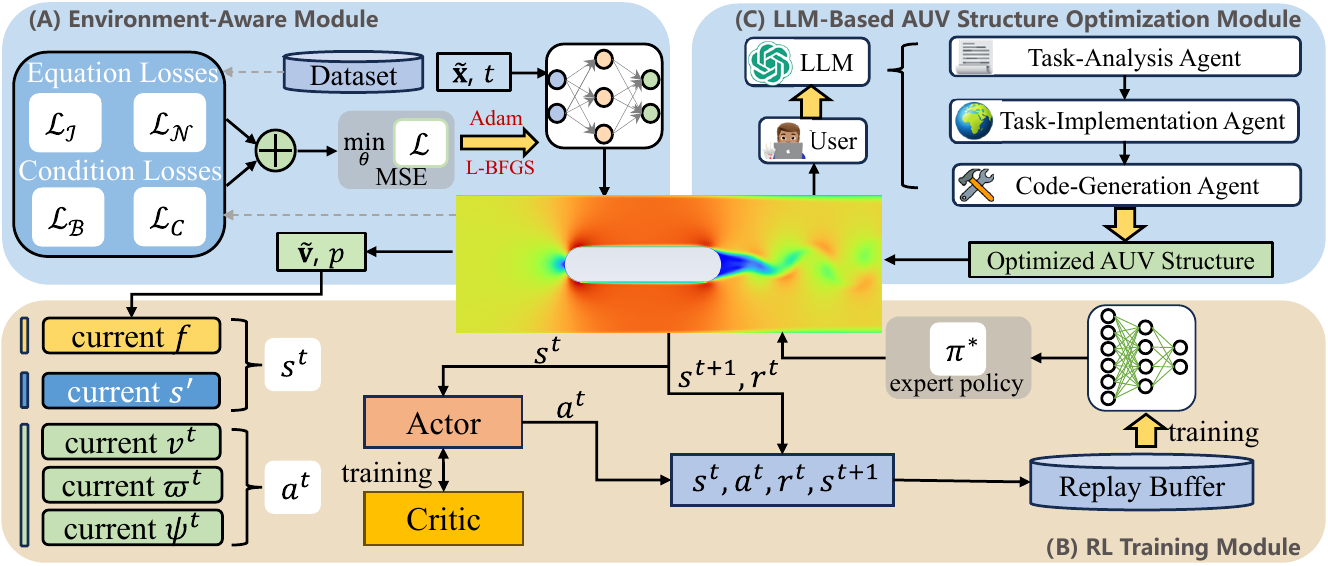}
\caption{\small \textbf{The overall architecture of our proposed environment-aware RL framework}, which comprises of three modules: (A) Environment-aware module. (B) RL-based training module. (C) LLM-based AUV structure optimization module.\vspace{-1.2em}}
\label{fig:2}
\end{figure*}

In this study, we employ the dimensionless Navier-Stokes equations and the continuity equation to characterize the underwater flow field. By transforming the problem of solving the system of equations into a constrained optimization problem, the environment-aware network can be trained to simulate the fluid dynamics equations. We define the losses for the Navier-Stokes equations, the continuity equation, the boundary conditions, and the initial conditions with their weight coefficients as $\alpha_1\mathcal{L}_{\mathcal{N}}(\mathbf{\vec{x}},t,\theta)$, $\alpha_2\mathcal{L}_\mathcal{C}(\mathbf{\vec{x}},t,\theta)$, $\alpha_3\mathcal{L}_\mathcal{B}(t,\theta)$, and $\alpha_4\mathcal{L}_\mathcal{I}(\mathbf{\vec{x}},\theta)$, respectively. The optimization problem is formulated as follows:
\begin{equation}
\begin{split}
OP\colon\min_{\theta}\mathcal{L}&=\alpha_1\mathcal{L}_{\mathcal{N}}(\mathbf{\vec{x}},t,\theta)+\alpha_2\mathcal{L}_\mathcal{C}(\mathbf{\vec{x}},t,\theta)+\alpha_3\mathcal{L}_\mathcal{B}(t,\theta)\\
&\quad +\alpha_4\mathcal{L}_\mathcal{I}(\mathbf{\vec{x}},\theta),
\end{split}
\end{equation} \vspace{-1.2em}
$s.t.$
\begin{equation}
\mathbf{\vec{x}\in[\mathbf{\vec{x}_{min}},\mathbf{\vec{x}_{max}}]},
\end{equation} \vspace{-1.2em}
\begin{equation}
 t\in[t_{min},t_{max}],
\end{equation} \vspace{-1em}
where
\begin{equation}
\begin{split}
    \mathcal{L}_{\mathcal{N}}(\mathbf{\vec{x}},t,\theta)=&(St\frac{\partial \mathcal{U}_E(\mathbf{\vec{x}},t;\theta)}{\partial t}+(\mathcal{U}_E(\mathbf{\vec{x}},t;\theta)\cdot\nabla)\mathcal{U}_E(\mathbf{\vec{x}},t;\theta)\\&+Eu\nabla p-\frac{1}{Re}\nabla^2\mathcal{U}_E(\mathbf{\vec{x}},t;\theta))^2,
\end{split}
\end{equation}  \vspace{-1.2em}
\begin{equation}
\begin{split}
    \mathcal{L}_{\mathcal{C}}(\mathbf{\vec{x}},t,\theta)=(\nabla\cdot\mathcal{U}_E(\mathbf{\vec{x}},t;\theta))^2,
\end{split}
\end{equation}  \vspace{-1.2em}
\begin{equation}
\begin{split}
    \mathcal{L}_{\mathcal{B}}(t,\theta)=(\mathcal{U}_E(\mathbf{\vec{x}}_\mathcal{B},t;\theta)-\mathcal{B}(\mathbf{\vec{x}}_\mathcal{B},t))^2,
\end{split}
\end{equation}  \vspace{-1.2em}
\begin{equation}
\begin{split}
    \mathcal{L}_{\mathcal{I}}(\mathbf{\vec{x}},\theta)=(\mathcal{U}_E(\mathbf{\vec{x}},t_\mathcal{I};\theta)-\mathcal{I}(\mathbf{\vec{x}},t_\mathcal{I}))^2,
\end{split}
\end{equation} 
where $St$, $Eu$ and $Re$ represent the Strouhal number, Euler number and Reynolds number, respectively. $\mathcal{B}(\mathbf{\vec{x}}_\mathcal{B},t)$ represents the boundary conditions at boundary $\mathbf{\vec{x}}_\mathcal{B}$, and $\mathcal{I}(\mathbf{\vec{x}},t_\mathcal{I})$ denotes the initial conditions at the initial time $t_\mathcal{I}$.


Parameter optimization via loss minimization allows the environment-aware module to accurately reconstruct flow fields. We use a hybrid optimization strategy that combines Adam and L-BFGS for training. Adam quickly reduces loss initially, while L-BFGS enhances parameter convergence later by leveraging second-order curvature information, striking a balance between computational efficiency and accuracy.

\subsection{The RL Training Module}
In traditional RL frameworks, RL training relies on two types of neural networks: the policy network and the value network. However, our environment-aware RL framework also integrates the environment-aware network, $\mathcal{U}_E(\mathbf{\vec{x}},t;\theta)$, which provides flow field information based on the current AUV structure. This flow field data is combined with other state information, such as distances between AUVs and target points, and incorporated into the state space. Consequently, during RL training, the AUV utilizes both flow field information and additional state data to enhance its decision-making and adapt more effectively to the underwater environment.

\textit{\textbf{Markov Decision Process Modeling}}: We model the environment-aware RL training process as a Markov Decision Process (MDP), which can be defined by a quintuple:
\begin{equation}\begin{split}
\mathcal{M}=&\{[{\mathcal{S}}_1^t,{\mathcal{S}}_2^t,\ldots,{\mathcal{S}}_N^t],[{\mathcal{A}}_1^t,{\mathcal{A}}_2^t,\ldots,{\mathcal{A}}_N^t],\\&[{\mathcal{R}}_1,{\mathcal{R}}_2,\ldots,{\mathcal{R}}_N],{\mathcal{P}}({s}^{t+1}|{s}^t, {{a}}^t),\gamma\},
\end{split}\end{equation}
where $N$ represents the number of AUVs required to complete the underwater task, ${\mathcal{S}}=[{\mathcal{S}}_1^t,{\mathcal{S}}_2^t,\ldots,{\mathcal{S}}_N^t]$ and ${\mathcal{A}}=[{\mathcal{A}}_1^t,{\mathcal{A}}_2^t,\ldots,{\mathcal{A}}_N^t]$ respectively represent the state space and action space at time $t$. ${\mathcal{R}=[{\mathcal{R}}_1,{\mathcal{R}}_2,\ldots,{\mathcal{R}}_N]}$ is the reward function, ${\mathcal{P}}(s^{t+1}|s^t, a^t) \in [0,1]$ indicates the state transition probability distribution, representing the probability of reaching state $s^{t+1}$ after state $s^t$ performs action $a^t$. $\gamma\in [0,1]$ stands for the discount factor. The detailed designs for each of these elements are described as follows:

\textbf{State Space $\boldsymbol{\mathcal{S}}$:} The state space in the environment-aware RL comprises two components: the AUV's observed information $s^{'}$, and features $f$ extracted from the flow field, represented as $\mathcal{S}=(s^{'},f)\in\boldsymbol{S^{'}}\times\boldsymbol{\mathcal{F}}$, where $\boldsymbol{S^{'}}$ and $\boldsymbol{\mathcal{F}}$ denote the state intervals observed by the AUV and the flow field information interval, respectively. By incorporating flow field data into the state space, the AUV's policy can adapt to varying flow conditions during training, facilitated by the environment-aware network based on PINN, as discussed in Section III-A. Additionally, we denote the AUV's position in the coordinate system as $\boldsymbol{\vartheta}^t=[p_x^t,p_y^t,p_z^t]^T$ to represent the position of the AUV in the coordinate system, where $(p_x^t,p_y^t,p_z^t)$ represent the AUV's horizontal, longitudinal, and depth positions, respectively.

\textbf{Action Space $\boldsymbol{\mathcal{A}}$:} In this study, we employ the three-degree-of-freedom underactuated model to describe the motion of the AUV. We use ${{a}}^t=[v^t,\varpi^t,\psi^t]^T$ according to current policy $\pi$ to characterize the AUV's action at time $t$, where $v^t$, $\varpi^t$ and $\psi^t$ respectively denote the AUV's expected velocity magnitude, yaw angle, and pitch angle at time $t$.

\textbf{Reward Function $\boldsymbol{\mathcal{R}}$:} In the MDP, the AUV evaluates action policies based on environmental rewards. During environment-aware RL training, as the AUV adapts to the underwater flow field, task completion improves compared to traditional RL frameworks, leading to higher rewards. The reward function typically includes components like failure penalties, efficiency promotion, and cooperation encouragement, with its design varying depending on the specific task.

\textit{\textbf{Remark.}} The hydrodynamic flow field around the AUV significantly influences its navigation resistance, with elevated resistance causing proportional increases in energy expenditure. Our RL framework integrates a multi-objective reward mechanism that simultaneously addresses collision prevention, task completion reliability, and energy consumption control. Beneficial flow conditions inherently reduce energy consumption, thereby enhancing reward signals to optimize navigation policies through fluid-environment awareness. This improvement is achieved by integrating hydrodynamic field characteristics into both state representation and reward calculation, enabling adaptive policy optimization that balances energy conservation with operational effectiveness.

Building on above analysis, further integration with RL algorithms (such as Soft Actor-Critic (SAC), Twin Delayed Deep Deterministic Policy Gradient (TD3), etc.) enables the optimization of the AUV's policy. Finally, AUV's expert policy $\pi^*$ possesses environment-aware capabilities, allowing it to adapt to the task scenario and the underwater flow field environment.


\renewcommand{\thefigure}{4}
\begin{figure*}[h]
	
	\begin{minipage}{0.315\linewidth}
		\centerline{\includegraphics[width=\textwidth]{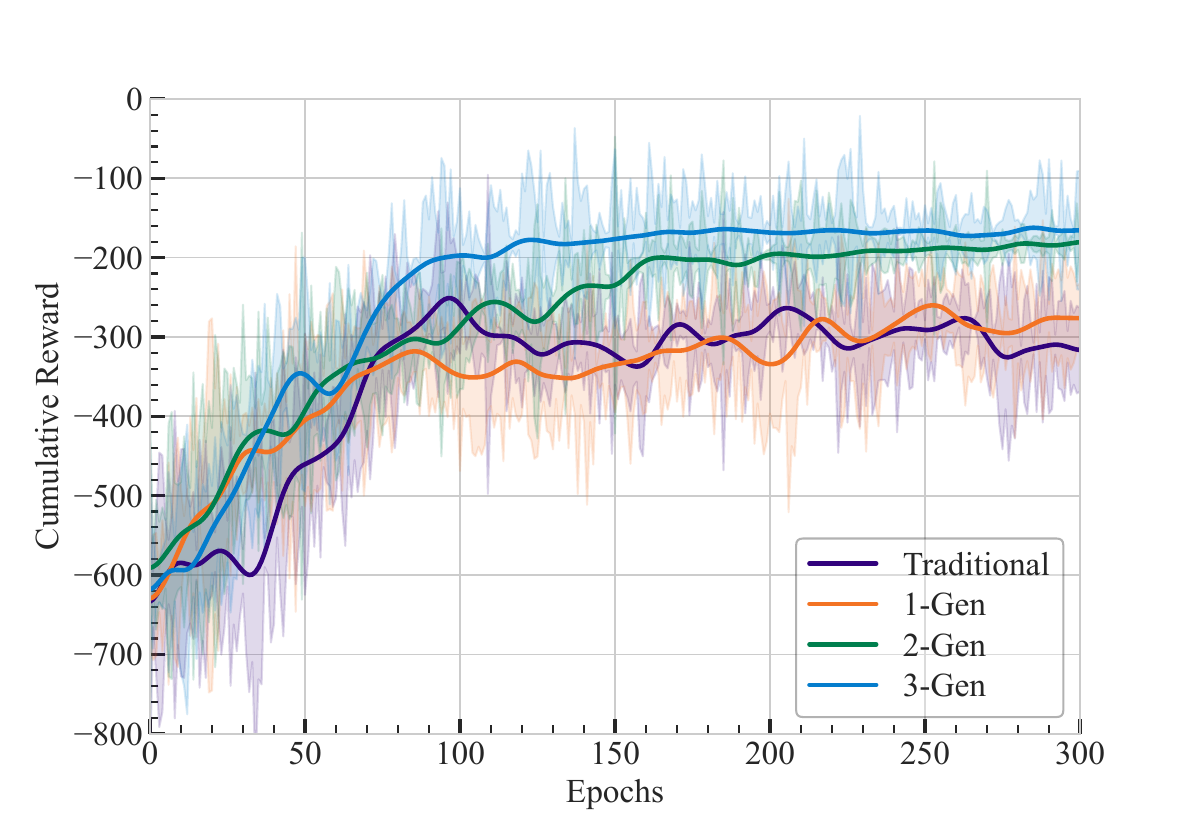}}
		\centerline{\small (a) Cumulative Reward (CR)}
	\end{minipage}
	\begin{minipage}{0.31\linewidth}
		\centerline{\includegraphics[width=\textwidth]{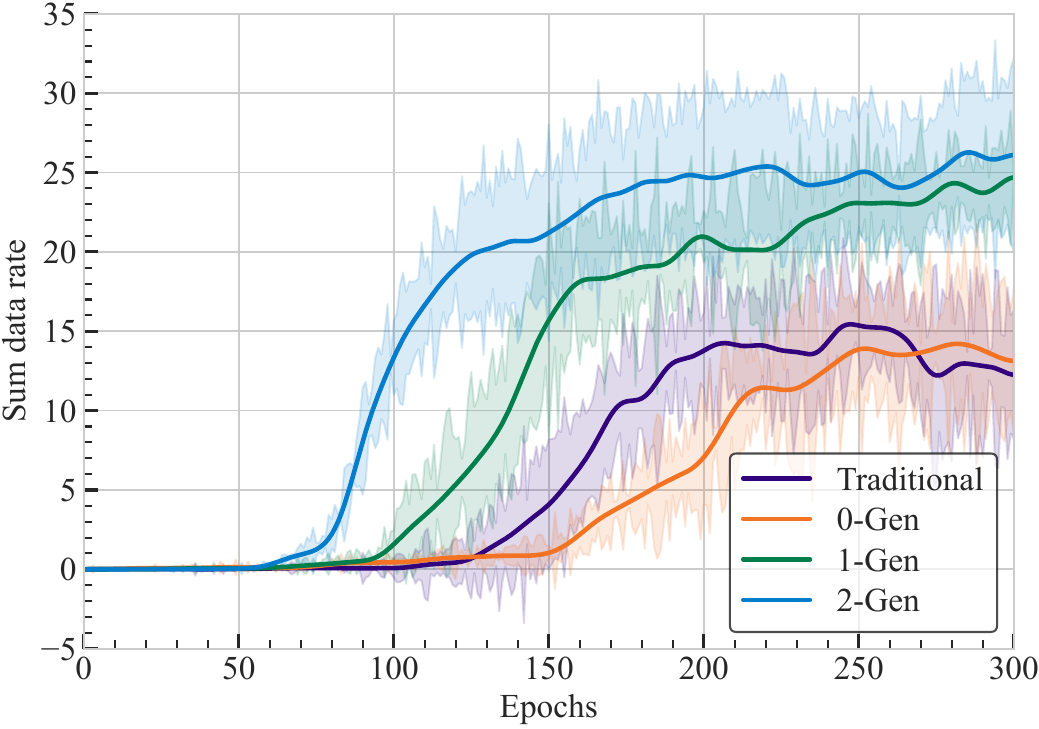}}
		\centerline{\small (b) Sum Data Rate (SDR)}
	\end{minipage}
	\begin{minipage}{0.315\linewidth}
		\centerline{\includegraphics[width=\textwidth]{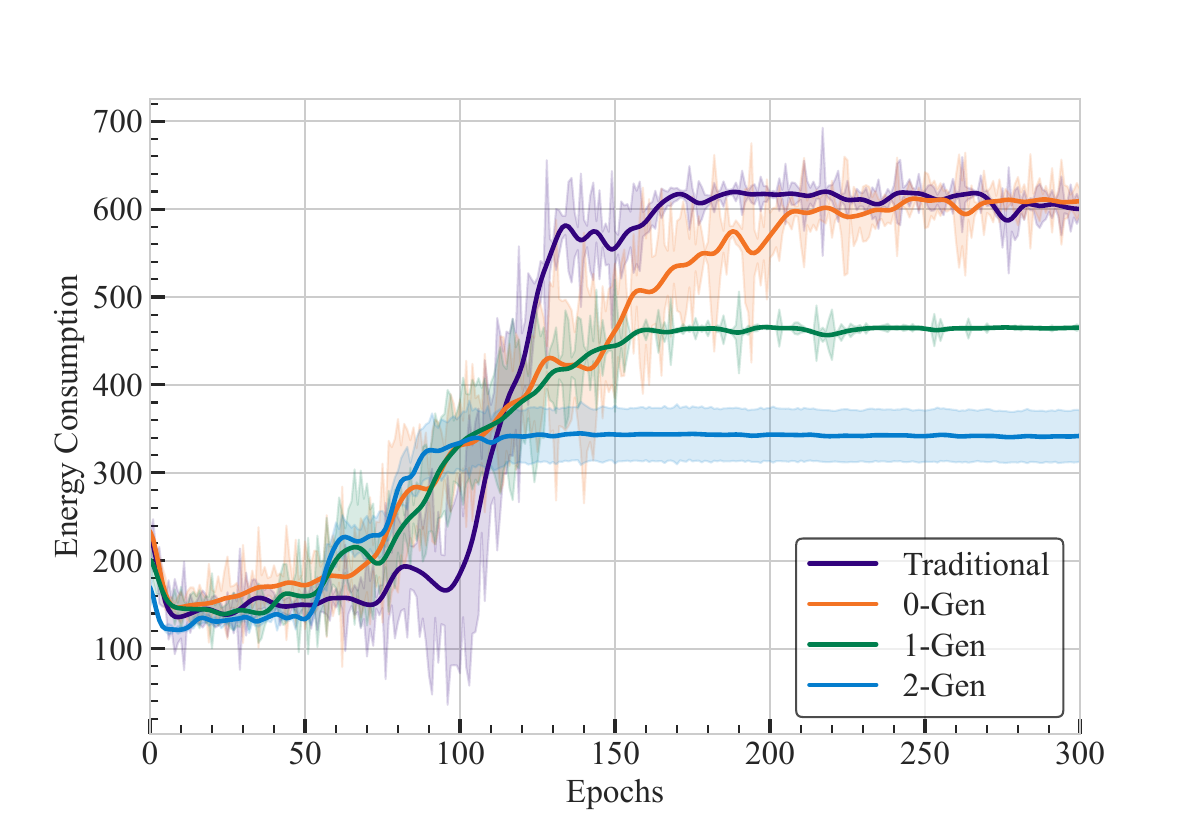}}
	 
		\centerline{\small (c) Energy Consumption (EC)}
	\end{minipage}
 
	\caption{\small \textbf{Performance Metrics Comparison: Three generations of AUVs under our proposed RL framework vs. traditional RL framework:} (a) Cumulative Reward (CR). (b) Sum Data Rate (SDR). (c) Energy Consumption (EC).\vspace{-1.2em}}
	\label{fig:4}
\end{figure*}

\vspace{-0.4em}
\subsection{LLM-Based AUV Structure Optimization Module}

LLM is trained on extensive text data and demonstrate strong problem-solving and content generation capabilities when provided with high-quality prompts. In our proposed environment-aware RL framework, we develop an efficient LLM-based AUV structure optimization module, which gradually refines the AUV's structure design based on task performance and flow field information around the AUV, making it more adaptive to complex underwater environment for accomplishing the required tasks. Thus, the objective of environment-aware RL can be transfered into a joint optimization of the AUV's structure (represented as flow field features $f$) and policy $\pi$, which is formulated as follows:
\begin{equation} \pi^*=
\underset{{f}}{\mathrm{argmax}} \underset{T \rightarrow \infty}{\mathrm{lim}} \mathbb{E}_{\pi} \left[\sum_{t = 0}^{T}\gamma^{t} \mathcal{{R}} \left(\pi, f\right)\right].
\end{equation}

\renewcommand{\thefigure}{3}
\begin{figure}
\centering
\includegraphics[width=0.928\linewidth]{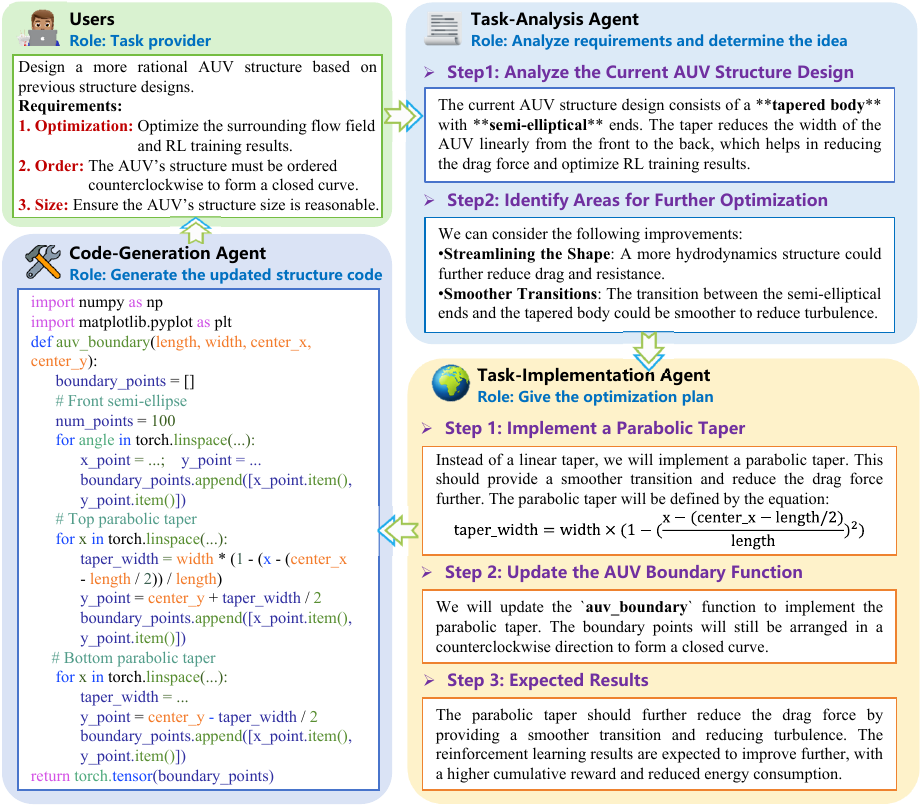}
\caption{\small \textbf{The workflow of the LLM-based AUV structure optimization module}, which demonstrates the interaction between the user and the task-analysis agent, task-implementation agent and code-generation agent, ultimately leading to the iterative optimization of the AUV structure design. \vspace{-2.6em}}
\label{fig:3}
\end{figure}
Fig. 3 illustrates the operational workflow of the LLM-based structure optimization module. Each task-specific agent, with domain expertise, begins the optimization process upon receiving assignments. This module comprises three collaborative agents: the Task-Analysis Agent, Task-Implementation Agent, and Code-Generation Agent. The workflow starts when \textbf{Users} submit task requirements and implementation goals, including the AUV's geometric configuration, flow field characteristics, RL training outcomes, and optimization objectives. This information is sent to the \textbf{Task-Analysis Agent}, which decomposes the task and proposes optimization strategies. The \textbf{Task-Implementation Agent} then initiates geometric modeling using mathematical formulations and computational geometry to create the optimized AUV structure. Following this, the \textbf{Code-Generation Agent} synthesizes Python simulation code based on the optimized design parameters. All design iterations are archived in a knowledge repository for continuous refinement. This architecture supports closed-loop feedback, ensuring that each agent's output is validated before proceeding, which maintains design coherence and computational efficiency.

\renewcommand{\thefigure}{5}
\begin{figure*}[h]
\centering
\includegraphics[width=0.868\linewidth]{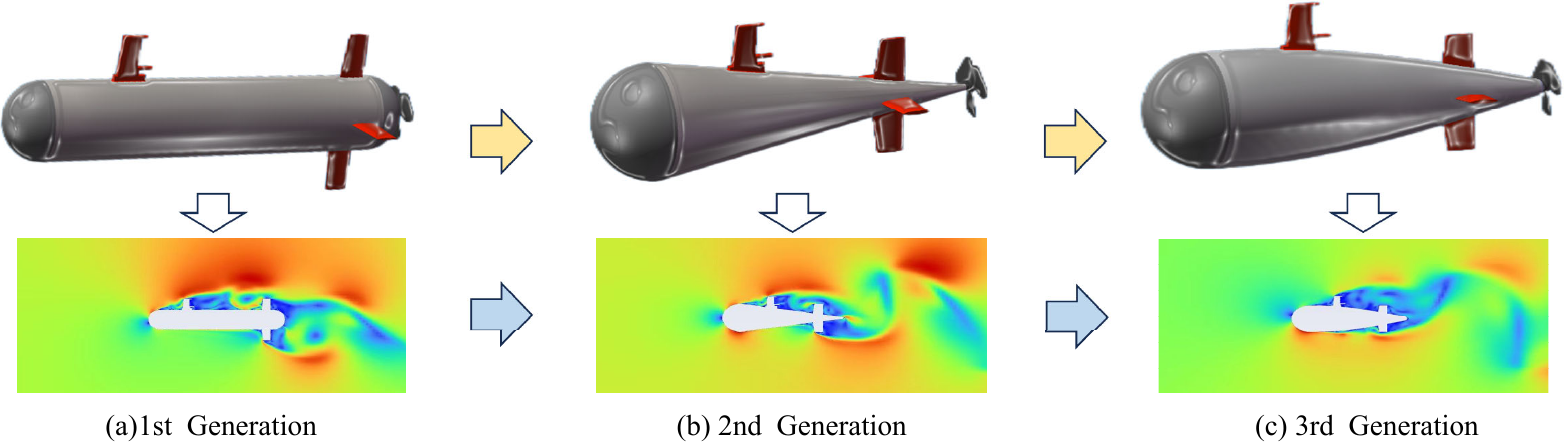}
\caption{\small \textbf{Schematic diagram illustrating the iterative evolution of three generations of AUV structural designs and their corresponding surrounding flow field environments:} (a) 1st Generation: Capsule-structured design; (b) 2nd Generation: Ice cream cone-structured configuration; (c) 3rd Generation: Teardrop-structured morphology. \vspace{-1.2em}}
\label{fig:5}
\end{figure*}

\begin{table}[!t]
\caption{Parameters of The Experiments and Algorithm.\label{tab:1}}
\centering
\setlength{\tabcolsep}{0.5mm}{
\begin{tabular}{m{4.7cm}<{\centering}c}
\hline
{\bf Parameters} & {\bf Values}\\
\hline
Experimental area & 200$\mathrm{m}$ $\times$200$\mathrm{m}$ $\times$200$\mathrm{m}$\\
Length of AUVs  & 1.5 $\mathrm{m}$\\
Width of AUVs  & 0.5 $\mathrm{m}$\\
Detection range  & 16.0 $\mathrm{m}$\\
Maximum speed  & 2.2 $\mathrm{m/s}$\\
Minimum speed & 1.2 $\mathrm{m/s}$\\
Pitch Angle range & [0,2$\pi$)\\
Yaw Angle range & [0,2$\pi$)\\
Crash distance & 5 $\mathrm{m}$\\
Training episodes & 400\\
The learning rate & 1$\times10^{-4}$\\
Replay buffer size & $2\times 10^4$\\     
\hline
\end{tabular}}
\end{table}

\section{EXPERIMENTS}
In this section, we first introduce our experiment setup in detail, then we conduct comprehensive simulation experiments to validate the feasibility and performance of our proposed environment-aware RL framework.

\subsection{Experiment Setup}

Key experimental parameters and configurations of our experiments are detailed in Table \uppercase\expandafter{\romannumeral1}. In these experiments, we employed the SAC RL algorithm with default settings \cite{37}. The experimental code was run on a personal computer with a 13th Gen Intel® Core™ i7-13650HX processor and NVIDIA GeForce RTX 4060 Laptop GPU. The superior performance of our framework was validated under the following scenarios:

• \textbf{Multi-AUV data collection task:} This task involves the coordinated operation of two AUVs in a 200 m$\times$200 m$\times$200 m area. to maximize data collection rates by prioritizing data from urgent nodes while minimizing energy consumption and avoiding collisions. Two experimental scenarios are established: normal and complex sea conditions, to assess the framework's robustness against environmental disturbances. For further details on this task, due to the constraint of space,
please refer to this previous work \cite{38}.

• \textbf{Multi-AUV target tracking task:} This task deploys multiple AUVs in coordinated multi-agent collaboration to pursue an underwater moving target. The AUVs continuously aware environmental informations, such as flow field and relative position, enabling dynamic motion adaptation through velocity/direction optimization for trajectory convergence with the target's path.

\subsection{Experiment I: Multi-AUV Data Collection Task Under Normal Sea Conditions}

We first applied environment-aware RL framework to train two AUVs to complete the data collection task under normal sea conditions, excluding complex disturbances like tides and turbulence. The AUVs utilized the embedded environment-aware module to capture near-field hydrodynamic data, which was integrated into the state space representation. After each training session, the LLM-based AUV structure optimization module quantitatively assessed the performance metrics in the training results and dynamically adjusted AUV morphology. Our experiment involved three iterative design generations, each consisting of 400 RL training episodes (results from the first 300 training episodes), whose results are shown in Fig. 4.

Fig. 4 presents a comparative analysis of training results of environment-aware RL versus the traditional RL framework. Results show that the environment-aware RL framework consistently outperforms the traditional RL framework across all AUV generations (notably in early performance gains and convergence speed). Both cumulative reward and data rate improved continuously during training, indicating the development of adaptive control policy for environmental compliance and task objectives. Additionally, energy consumption stabilized quickly despite performance gains, demonstrating the proposed framework's energy efficiency. Evolving structure designs yielded significant improvements in reward and data rate while reducing energy consumption, suggesting that optimized AUV's structure enhance environmental adaptability through reduced drag and improved flow field. The details of AUV structure and the structure of each AUV generation with corresponding flow field information are presented in Table \uppercase\expandafter{\romannumeral2} and Fig. 5, respectively.

Fig. 5 depict the design refinements across three AUV generations, along with their corresponding flow field environments. As a schematic illustration, Fig. 5 presents flow fields around precise geometric configurations of three generations of AUV, simulated via transient Navier-Stokes equations using ANSYS Fluent. The initial capsule-structured design offers stability in underwater navigation but is suboptimal for complex tasks. Using RL training results and flow field data from the 1st generation (1-gen), the 2nd generation (2-gen) AUV was redesigned with an "ice cream cone" structure, improving flow distribution and reducing drag while enhancing maneuverability and stability. While the 2nd generation features a linear conical structure, the 3rd generation (3-gen) adopts a parabolic tail for further drag reduction. Additionally, the transition between the semicircular head and conical body in the 3rd generation is smoother, minimizing turbulence and enhancing the AUV's efficiency and task completion capabilities. Table \uppercase\expandafter{\romannumeral2} compares the structural designs of three generations of AUVs, where $\zeta=\frac{x-(C_x-L/2)}{L}$, $C_x$ and $C_y$ denote the horizontal and vertical coordinate of the central point, and $W$ and $L$ represent the width and length of the AUV, respectively.


\begin{table}[t]
    \centering
    \caption{Comparison of three-generation of AUV shapes.}
    \label{table:masking_performance}
    \setlength{\tabcolsep}{4pt}
    \begin{tabular}{lcccc}
    \toprule
      & Capsule Structure & Ice Cream Shape & Parabolic Tail  \\
    \midrule
    Equation  & $C_y\pm \frac{W}{2}$ & $C_y\pm \frac{W(1-\zeta )}{2}$ & $C_y\pm \frac{W(1-\zeta^2)}{2}$ \\
    Curvature & Abrupt & Constant  & Gradient  \\
    Flow Field  & Turbulence prone   & Controlled separation  & Low Resistance  \\
    \bottomrule
    \end{tabular}
\end{table}

	

\renewcommand{\thefigure}{6}
\begin{figure}[ht]
	
	\begin{minipage}{0.489\linewidth}
		\centerline{\includegraphics[width=\textwidth]{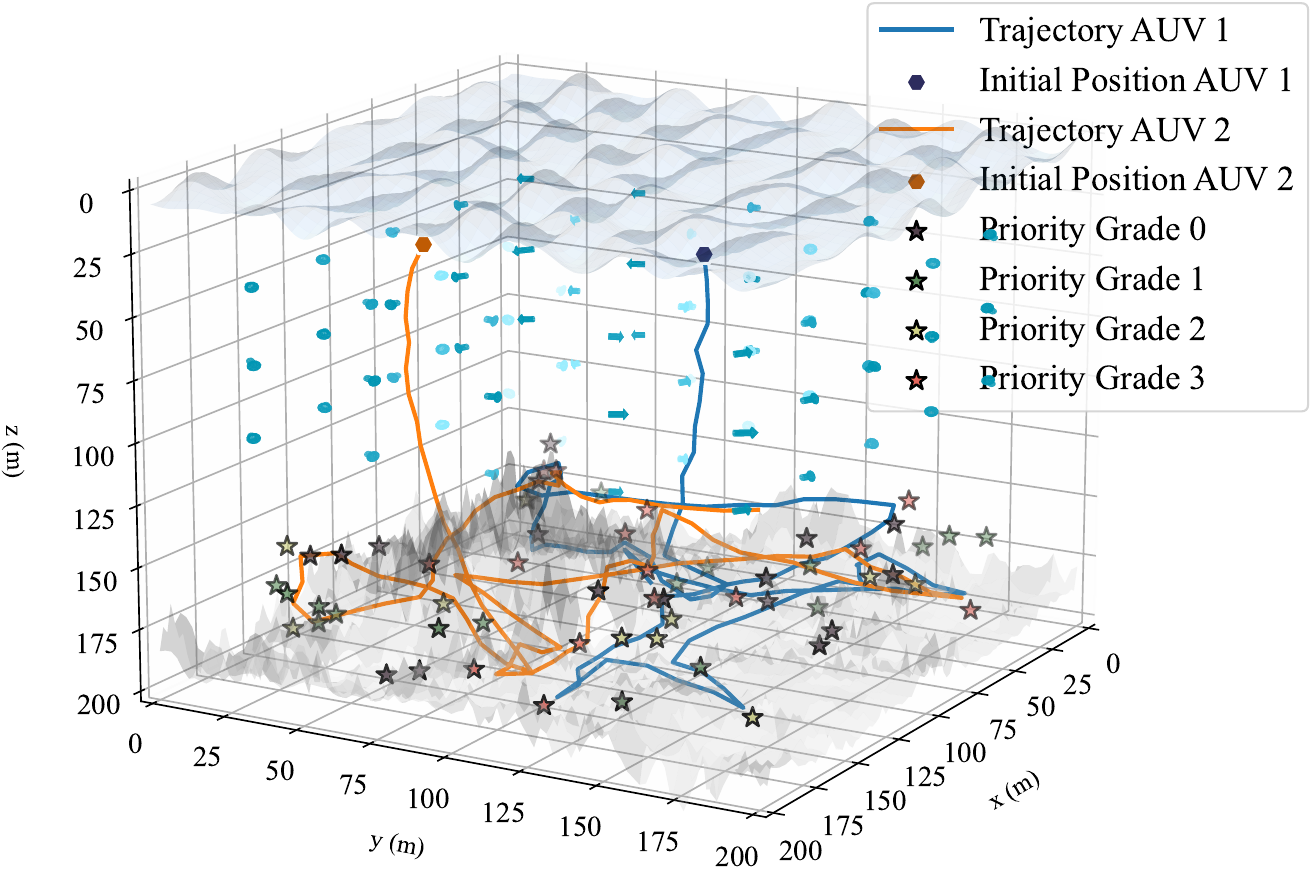}}
		\centerline{\small (a) Env-aware RL framework}
	\end{minipage}
	\begin{minipage}{0.489\linewidth}
		\centerline{\includegraphics[width=\textwidth]{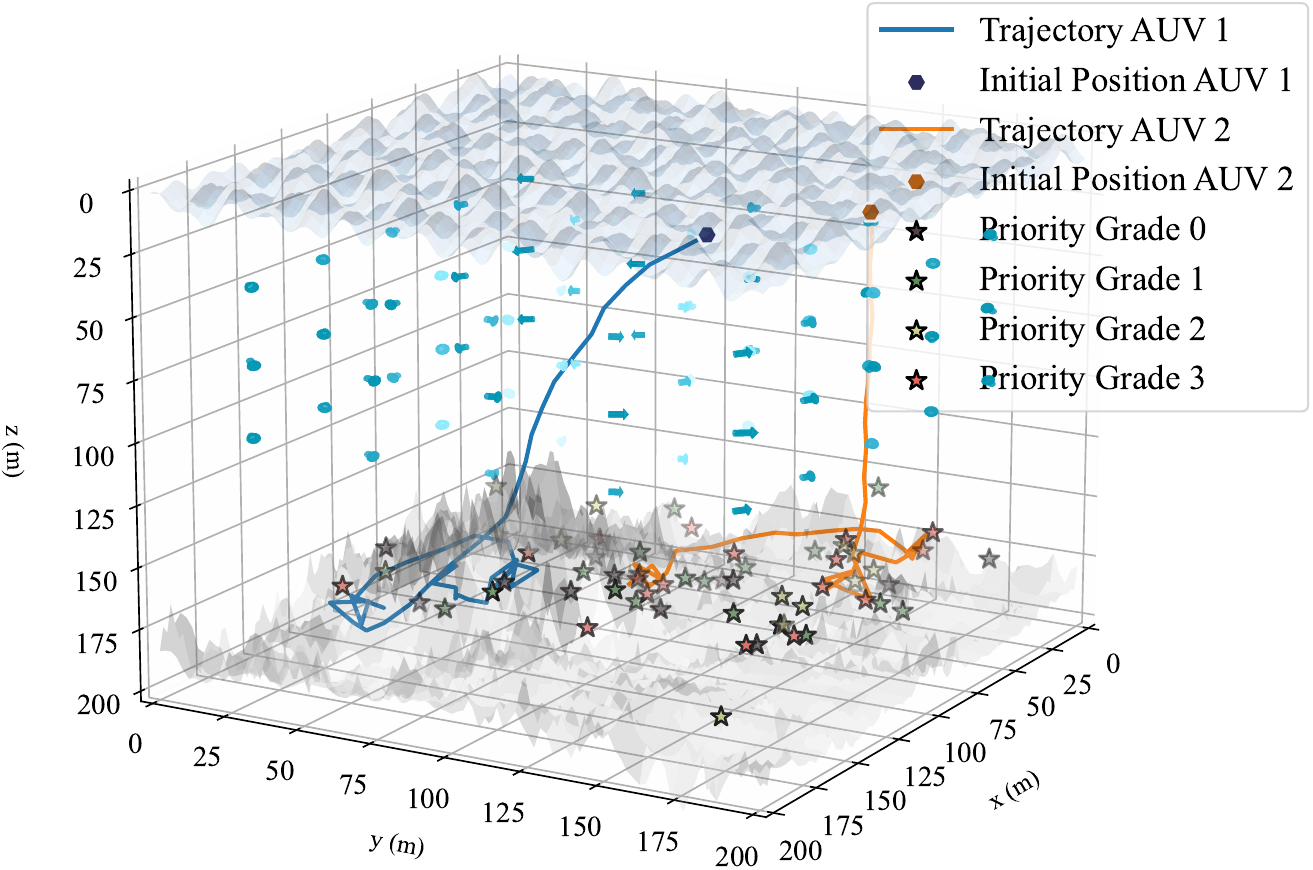}}
		\centerline{\small (b) Traditional RL framework}
	\end{minipage}
	\caption{\small \textbf{Performance Comparison: The trajectories of AUVs under complex sea conditions trained through the environment-aware RL framework. vs. traditional RL framework:} (a) The environment-aware RL framework. (b) The traditional RL framework.}
	\label{fig:7}
\end{figure}

\renewcommand{\thefigure}{7}
\begin{figure*}[ht]
	
	\begin{minipage}{0.33\linewidth}
		\centerline{\includegraphics[width=\textwidth]{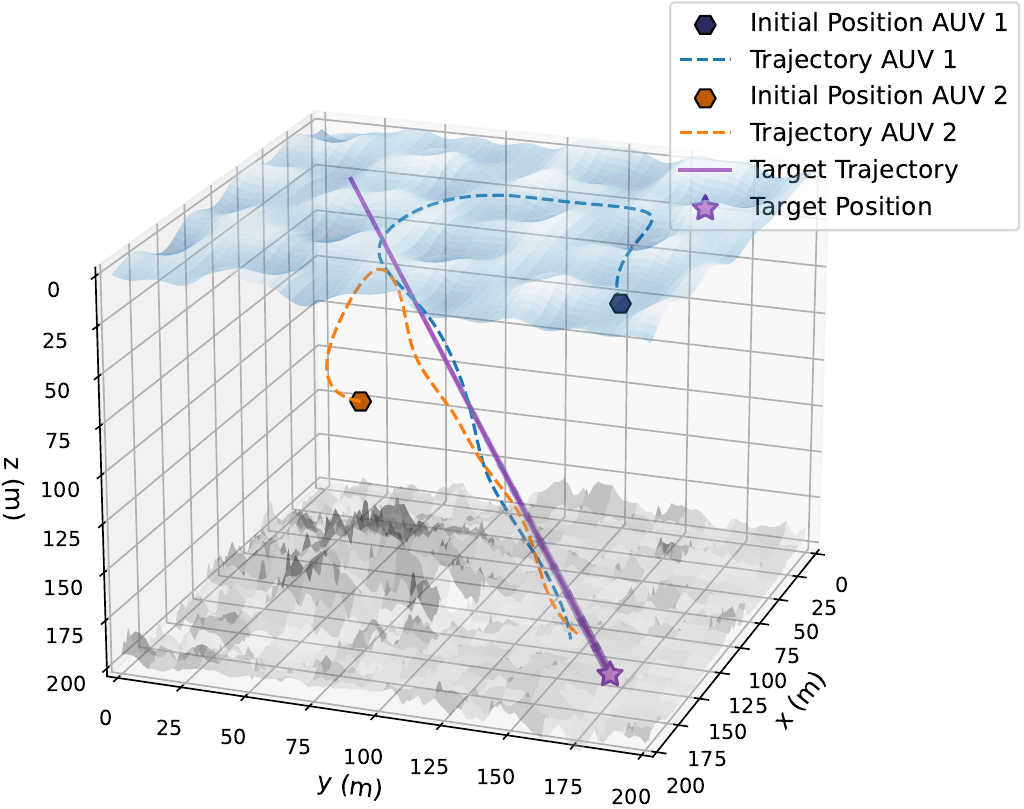}}
		\centerline{\small (a) Straight Line}
	\end{minipage}
	\begin{minipage}{0.32\linewidth}
		\centerline{\includegraphics[width=\textwidth]{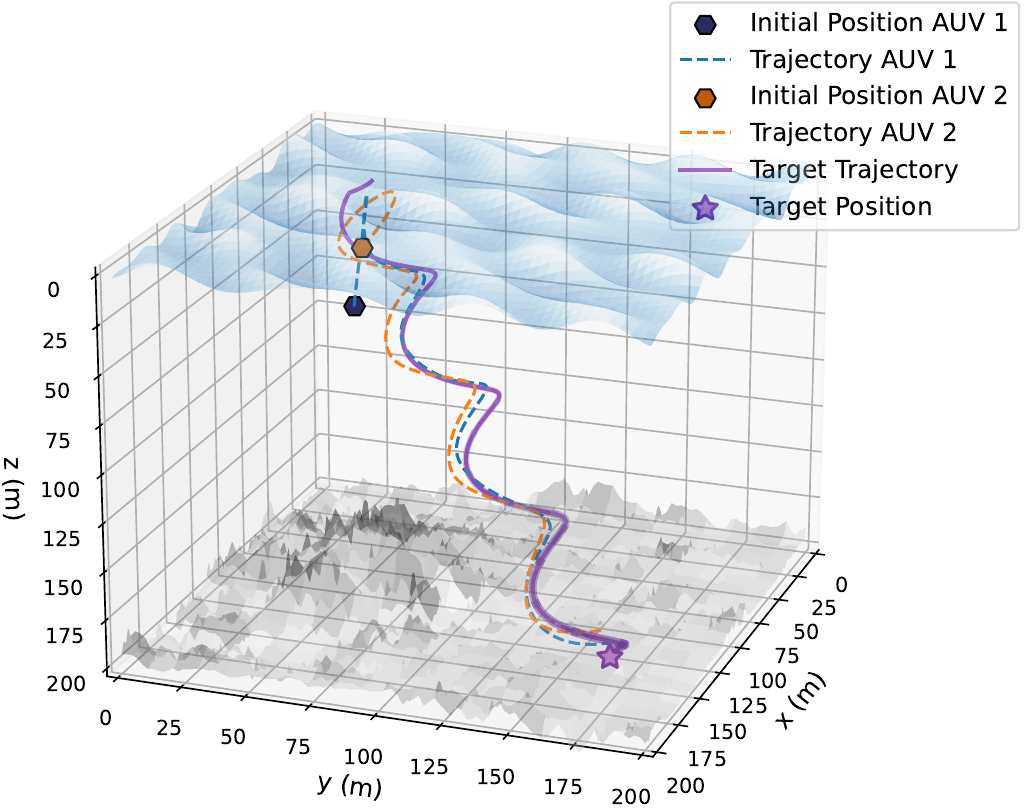}}
		\centerline{\small (b) Sinusoid}
	\end{minipage}
	\begin{minipage}{0.325\linewidth}
		\centerline{\includegraphics[width=\textwidth]{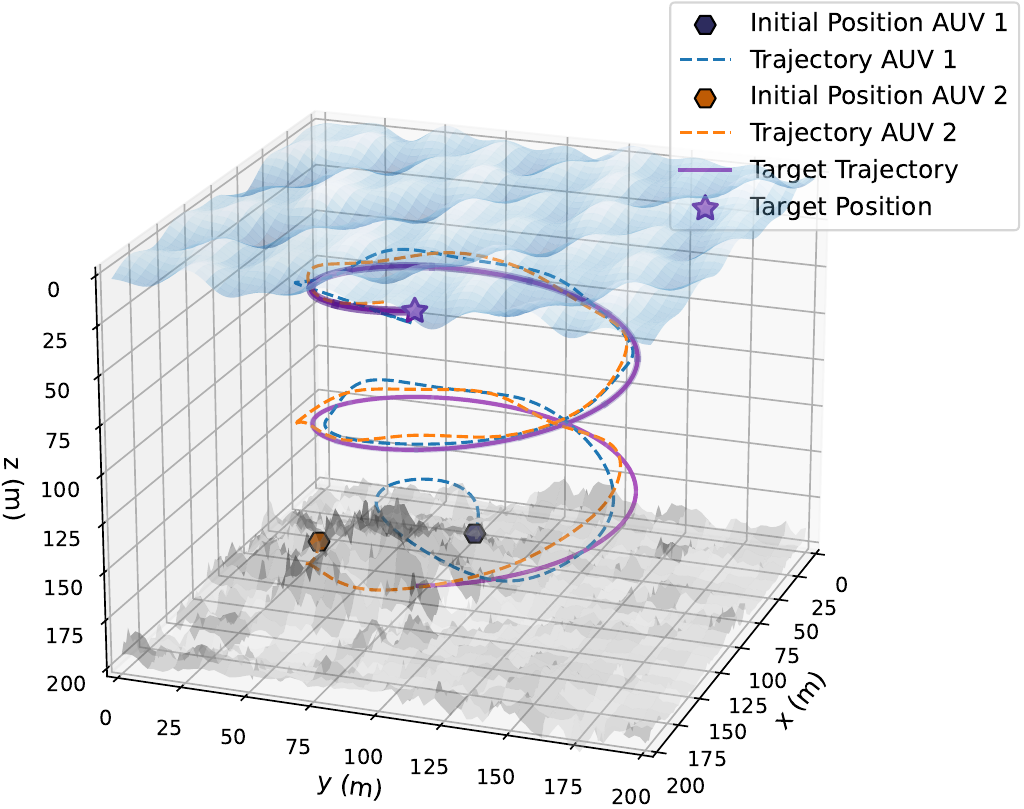}}
	 
		\centerline{\small (c) Spiral Line}
	\end{minipage}
 
	\caption{\small \textbf{Trajectory diagrams for multi-AUV target point tracking task under three scenarios:} (a) Straight Line. (b) Sinusoid. (c) Spiral Line.}
	\label{fig:8}
\end{figure*}

\subsection{Experiment II: Multi-AUV Data Collection Task Under Complex Sea Conditions}

To assess the robustness of our proposed framework in challenging conditions, consider an AUV system facing resistance from complex currents and tidal waves. Surface fluctuations can cause positional disturbances, while underwater currents may alter AUV motion. We simulate these conditions using shallow water equations and second-order Stokes waves. Two AUVs are deployed to perform data collection in harsh underwater environments.

Fig. 6 presents a comparative analysis of motion trajectories during underwater operations. Subfig. (a) shows that AUVs trained with the environment-aware RL framework successfully meet data collection requirements by prioritizing high-priority nodes, avoiding collisions, and maximizing coverage, thus improving the overall data rate. They also minimize redundant trajectories to control energy consumption. The comparison reveals that our framework significantly outperforms the traditional RL framework in both data collection coverage and task completion metrics, demonstrating enhanced adaptability and execution capabilities. This trajectory comparison supports the training data results in Fig. 4, confirming the effectiveness of our framework.

To validate the generalizability and superior performance of the proposed framework, we integrate SAC and TD3 into the environment-aware RL framework and compare them with traditional RL across key metrics. As shown in Table \uppercase\expandafter{\romannumeral3}, the environment-aware SAC achieves a maximum sum data rate (SDR) of 33.12 and a cumulative reward (CR) of -138, while the environment-aware TD3 attains a SDR of 35.94 and a CR of -149, significantly outperforming the traditional RL's CR of -180 and SDR of 25.82. Notably, both algorithms integrated into the environment-aware framework exhibit markedly reduced energy consumption (EC), with SAC and TD3 converging at 340.44 and 342.56, respectively, contrasting sharply with the traditional RL’s substantially higher EC of 610.49. These findings validate the broad applicability of the environment-aware RL framework and its efficacy in balancing task efficiency, operational reliability, and energy sustainability.

\begin{table}[t]
    \centering
    \caption{Data collection performance metrics comparison.}
    \label{table:masking_performance2}
    \setlength{\tabcolsep}{4pt}
    \begin{tabular}{lcccc}
    \toprule
      & Env-aware SAC & Env-aware TD3 & Traditional RL  \\
    \midrule
    CR  & -138 & -149 & -180 \\
    SDR & 33.12 & 35.94  & 25.82  \\
    EC  & 340.44   & 342.56  & 610.49  \\
    \bottomrule
    \end{tabular}
\end{table}


\subsection{Experiment III: Multi-AUV Target Tracking Task}

To validate the generalizability of environment-aware RL framework for more underwater tasks, we conducted an effectiveness verification using a multi-AUV target tracking task. In this experiment, two AUVs were deployed in a marine environment to perform real-time trajectory tracking of a moving target with three distinct motion patterns. We utilized three iterative design generations, each with 300 RL training episodes, to rigorously evaluate adaptive navigation capabilities under varying hydrodynamic conditions.

The results of the target tracking experiment are shown in Fig. 7. Both AUVs successfully converged on the target's trajectory, demonstrating effective real-time tracking performance. Even when starting from random positions away from the target path, the AUVs made responsive navigation adjustments through coordinated heading changes and velocity regulation. Systematic observations indicate that both platforms consistently aligned their trajectories quickly, adapting dynamically to the moving target. These outcomes validate the framework's robust adaptability and reliable task execution in diverse operational environments.

In this study, the trajectory tracking success rate is defined as the ratio of target entries into the AUV's detection range to total movement actions per RL training episode, serving as a quantitative metric for tracking performance. Table \uppercase\expandafter{\romannumeral4} compares individual and combined tracking success rates (i.e., simultaneous target acquisition by both AUVs) across different algorithms. AUVs trained with our environment-aware RL framework achieved success rates exceeding 85\%, demonstrating statistically significant superiority over conventional RL approaches. This conclusively validates the enhanced efficacy of the proposed framework.

\begin{table}[t]
    \centering
    \caption{Trajectory tracking success rate comparison.}
    \label{table:masking_performance3}
    \setlength{\tabcolsep}{4pt}
    \begin{tabular}{lcccc}
    \toprule
      & Env-aware SAC & Env-aware TD3 & Traditional RL  \\
    \midrule
    AUV1  & 98.2\% & 85.8\% & 76.4\% \\
    AUV2 & 96.3\% & 97.1\%  & 82.4\%  \\
    Simultaneously  & 85.8\%   & 89.0\%  & 69.8\%  \\
    \bottomrule
    \end{tabular}
\end{table}

\section{CONCLUSION}

This paper proposes an environment-aware RL framework to improve AUV's performance in underwater tasks. Extending traditional RL architecture, we introduce an environment-aware network module that integrates flow field data into the state space for real-time environmental adaptation. The framework also optimizes AUV structural performance through LLM-based iterative refinement, leveraging environmental conditions and training feedback to enhance operational capabilities. Experimental validation demonstrates the framework's effectiveness and robustness.

\section{ACKNOWLEDGEMENT}
The authors gratefully acknowledge the anonymous reviewers for their constructive comments. We also extend our sincere thanks to Dr. Xiangwang Hou, Prof. Yong Ren, Prof. Daoyi Chen, and Prof. Juntian Qu from Tsinghua University for their insightful discussions and guidance. Additionally, we appreciate the encouragement and recognition from Prof. Nare Karapetyan at the Woods Hole Oceanographic Institution, Prof. Xiaofan Li at the University of Hong Kong, and Prof. Xiaomin Lin at the University of South Florida.





\bibliographystyle{IEEEtran}
\bibliography{IEEEexample}

\end{document}